# Screen Parsing: Towards Reverse Engineering of UI Models from Screenshots

Jason Wu*
HCI Institute, Carnegie Mellon University
Pittsburgh, PA
jsonwu@cmu.edu

Xiaoyi Zhang, Jeffrey Nichols, Jeffrey P. Bigham
Apple
Cupertino, CA
{xiaoyiz,jwnichols,jbigham}@apple.com

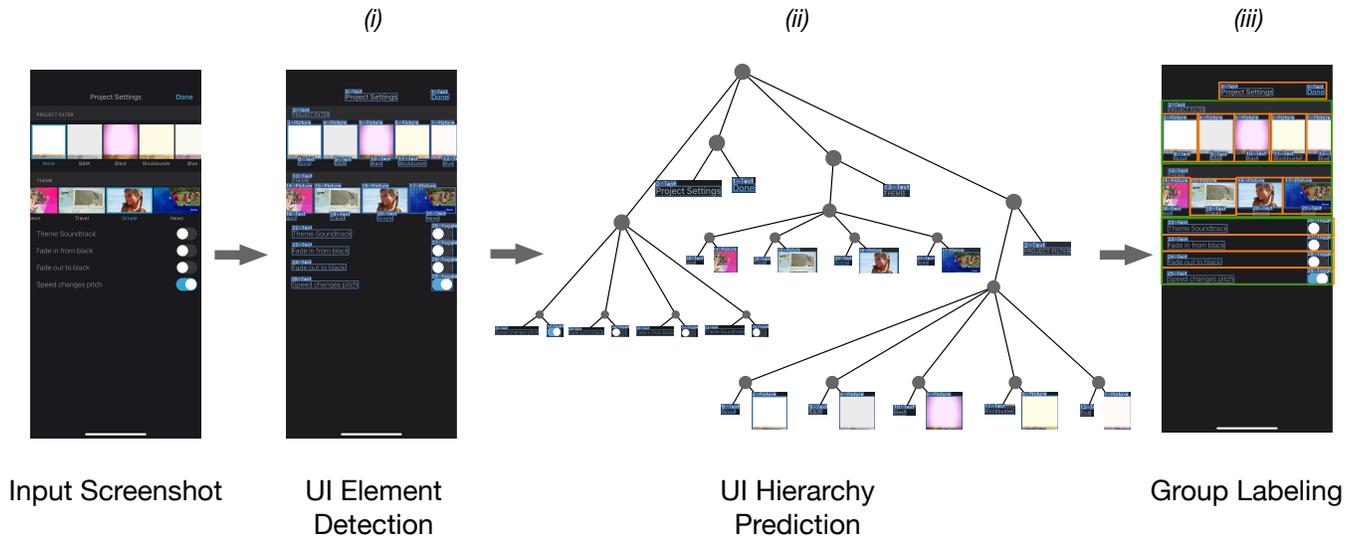

Figure 1: An overview of our implementation of *screen parsing*. To infer the structure of an app screen, our system *(i)* detects the location and type of UI elements from a screenshot, *(ii)* predicts a graph structure that describes the relationships between UI elements, and *(iii)* classifies groups of UI elements.

## ABSTRACT
Automated understanding of user interfaces (UIs) from their pixels can improve accessibility, enable task automation, and facilitate interface design without relying on developers to comprehensively provide metadata. A first step is to infer what UI elements exist on a screen, but current approaches are limited in how they infer how those elements are semantically grouped into structured interface definitions. In this paper, we motivate the problem of *screen parsing*, the task of predicting UI elements and their relationships from a screenshot. We describe our implementation of screen parsing and provide an effective training procedure that optimizes its performance. In an evaluation comparing the accuracy of the generated output, we find that our implementation significantly outperforms current systems (up to 23%). Finally, we show three example applications that are facilitated by screen parsing: *(i)* UI similarity search, *(ii)* accessibility enhancement, and *(iii)* code generation from UI screenshots.

## 1 INTRODUCTION
User interfaces are, unsurprisingly, designed for consumption by human beings, and it can be difficult for automated systems to understand what functionality is present in a user interface, how the different components of the interface work together, and how it can be operated to accomplish some goal. This is particularly true if the automated system does not have access to any meta-data about the user interface, such as view hierarchies or accessibility tags, or if this information is missing or incompletely defined, as is often the case. Automated user interface understanding systems could offer many benefits. For example, a screen reader (*e.g.,* VoiceOver and TalkBack) could facilitate access to user interfaces for blind and visually impaired users when the underlying app does not provide appropriate meta-data [41], and task automation agents (*e.g.,* Siri Shortcuts and IFTTT) could allow users to automate repetitive or complex tasks with their devices more efficiently. These benefits are gated on how well these systems can understand and interact with the underlying applications. Many of today's systems rely on the availability of UI meta-data and fail when this information is unavailable. To overcome this recent efforts have focused on predicting the presence of an app's on-screen elements solely from its visual appearance.

---

*This work was done while Jason Wu was an intern at Apple.

Structure is a core property of UIs that is reflected both in how they are constructed and how they are used. However, many current approaches to visual modeling of UIs ignore or fail to centralize this aspect. In this paper, we present a new approach called *screen parsing*, which applies techniques used in NLP for natural language parsing to produce machine-learned models that predict the UI hierarchy of an app from its screenshot. Our approach involves *(i)* a Faster-RCNN model for detecting the set of elements on a screen, *(ii)* a stack-based transition parser model for predicting the hierarchy of how those elements relate to each other, and *(iii)* a Deep Averaging Network model that classifies element groupings. We describe the details and training procedure of our implementation of screen parsing, and conduct an evaluation in which we compare the performance of our system against baseline approaches. Using a set of 5 metrics, we show that our implementation performs up to 23% better than baseline systems depending on the performance metric used. Finally, we show three example applications enabled by our implementation of screen parsing.

More broadly, we believe that systems can benefit from perceiving UI screens as humans do – not as a set of elements, but as a coordinated and organized presentation of content. Structural understanding is an important step that can help systems reason about relationships between interaction controls and content. Our model implementation is trained to predict one type of relation (links in the view hierarchy), but we believe screen parsing and our modeling approach can be extended to others as well (*e.g.,* navigation order).

To summarize, this paper makes the following contributions:

- A problem definition of *screen parsing* which is useful for a wide range of UI modeling applications.
- A description of our implementation of screen parsing and its training procedure.
- A comprehensive evaluation of our implementation with baseline comparison.
- Three implemented examples of how our model can be used to facilitate downstream applications such as *(i)* UI similarity, *(ii)* accessibility metadata generation, and *(iii)* code generation.

## 2 RELATED WORK
### 2.1 Reverse Engineering UIs
Many approaches to visual UI modeling focus on "reverse engineering" hidden attributes and potentially modifying them at runtime. Reverse-engineering methods often focus on extracting semantic attributes from visible information presented by the app (*i.e.,* pixel information), which allows them to support a broader array of use-cases.

An important use-case is facilitating non-visual access to apps for people with disabilities. Outspoken [34] was of one the first screen readers that supported GUIs, which required it to describe both text and graphical elements of the screen. To process icons and other pictorial elements, the system maintained a database of graphical elements (paired with a verbal description) and matched on-screen elements to descriptions of similar items. Today, the ecosystem of UI toolkits is much larger and permits much greater functionality, including allowing developers to embed icon and image descriptions in an app's metadata, yet many apps are still inaccessible because they do not include this data. To support inaccessible apps, recent screen reader technology [7, 41] uses deep convolutional neural networks to generate element descriptions and other accessibility metadata.

Reverse engineering methods can also be used to extend existing GUI applications. A common approach to interface with applications without an application programming interface (API) is to define "macros" that automate sequences of key-strokes and mouse movements. To acquire interaction targets, many automation toolkits provide functions for searching the screen for pixel values and returning their coordinates [2]. Sikuli [39] and PAX [5] are systems that improve the localization of targets by supporting more advanced matching techniques (*e.g.,* bitmap matching and heuristics) and combining hierarchical information extracted from an external source, such as the system window manager. Elements localized using pixel-based methods can also be used to modify apps at runtime [9], and previous work has investigated the benefit of hierarchy prediction (using heuristics) for this use-case.

Finally, reverse-engineering approaches have been applied to generate code from UI mockups or screenshots. A subset of these approaches have focused on translating hand-drawn wireframes to GUI code. These tools [1, 23] are useful for designers who wish to quickly sketch and prototype possible UI layouts. A more complex version of this task is generating code from complete UI screenshots, as it requires that the system handle the stylistic and structural variation present in real-world app screens. REMAUI [29] is a system that uses heuristics to combine OCR detection results and cropped patches from the original screenshot to generate working UI code. Pix2Code [4] is an end-to-end code generation model that uses a CNN encoder to encode a screenshot and a RNN decoder to generate code. UI2Code uses a similar architecture to generate a "GUI Skeleton" from a screenshot [6] that describes the relative positioning of UI elements.

### 2.2 Defining and Extracting UI Models
While reverse-engineering systems can effectively predict a subset of a screen's attributes, automated systems aimed at deeper and more complex interactions with UIs must support higher-level, semantic understanding of UIs. We reviewed literature related to model-based user interface (MBUI) development, which here we use as a conceptual framework for describing how UIs are constructed, presented, and used.

MBUI development refers to a development process that *(i)* first defines high-level models for an interface, then *(ii)* produces code that conforms to that model [14]. Models, among other things, detail what data the UI will display and how it will be used, and are a helpful tool for organizing the creation of UI applications. Puerta [32] describes an example of how this process is applied and categorizes common types of models (*e.g.,* data model, domain model, presentation model). Because a well-designed model can describe all or nearly all aspects of an interface, it is often possible to automatically generate code from model specifications [12, 30].

Similarly, it may be useful for automated systems to extract or infer models from a finished application, as doing so would reveal semantics. In this paper, we present an system that predicts the UI hierarchy (closely related to the *presentation model*) of an app from its screenshot. More broadly, our formulation of *screen parsing* as visual inference of structured relationships is useful for extracting UI models, which are often structured relationships among items.

## 2.3 Structured Prediction from Visual Information

To provide additional background about our work and opportunities for UI modeling, we review some machine learning approaches that can be used to predict structure from visual information.

Many approaches to structured prediction have their roots in natural language processing (NLP). Early work on scene segmentation used stochastic grammars to analyze layouts (known as geometric parsing [35]) or construct hierarchical representations from proposal regions (*i.e.,* image patches) [37, 42]. However, it can be difficult to define or induce a grammar that explicitly describes all primitives and relationships and work well with continuous attributes. Moreover, many grammars are designed to work with sequential input (common in language) rather than spatial input (common in vision). Socher et al. propose a more general architecture that learns to recursively join related items in both images and text using a neural network model [36].

More recent work in the computer vision literature has focused on visual scene understanding through *scene graphs*. Scene graphs represent relationships between objects detected in an image and are described as a collection of relationship triplets (<subject, predicate, object>) [20]. Approaches to scene graph detection vary – some models first perform object detection then consider all possible pairs [40], while others directly generate a set of likely relationships [38].

As we will discuss later, *screen parsing* is closely related to these structured visual understanding tasks and is targeted towards aspects UI modeling. The design of our model is also based on many of the same core ideas, which we implement in service of our task definition.

## 3 SCREEN PARSING
## 3.1 Problem Formulation

We define the problem of *screen parsing*, which we use to refer to the prediction of structured UI models from visual information. As a review, we use *UI models* to refer to high-level abstractions of UI semantics *e.g.,* logic, presentation, and associated tasks [14]. A *screen parsing* model takes a UI screenshot as input and produces a graph representation of a model as output. The connections in the graph can be used to express a variety of semantic and syntactic concepts. For example, one might use an edge to represent interaction flow (*e.g.,* the "Username" text field should be filled out before tapping on the "Login" button).

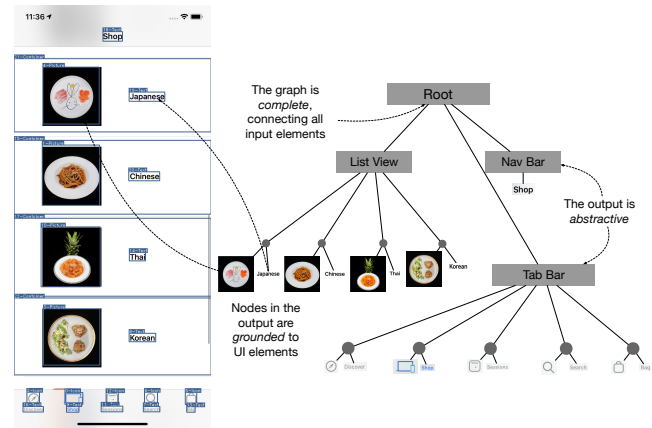

**Figure 2: We show an example of an input screen (Left) and the corresponding *screen parse* (Right). The graph contains all of the visible elements on the screen (the output is *complete*), groups them together to form higher-level structures (*abstractive*), and nodes can be used to reference UI elements (the output is *grounded*).**

In this paper, we focus on generating an app's UI hierarchy (*i.e.,* presentation model) which is a specification of how UI elements are grouped and rendered on the screen [32]. Figure 2 shows an example of a screen and corresponding UI hierarchy graph. The properties of UI hierarchies introduces some constraints on the types of valid outputs.

- *Complete* – the output is a single tree that spans all of the detected UI elements.
- *Grounded* – Nodes in the output reference specific UI elements in the screen.
- *Abstractive* – The output can group elements together (potentially more than once) to form higher-level structures.

Moreover, all UI hierarchies can be described as *directed trees*, which we constrain our system to generating.

## 3.2 Comparison to Related Problems

*Screen parsing* is closely related to and, in many ways, motivated by other problems in the UI modeling and computer vision. Specifically, we select three similar tasks for comparison: *(i)* UI Element Detection, *(ii)* GUI Skeleton Generation, and *(iii)* Scene Graph Generation. All of these approaches generate semantic output from a visual representation of a screen (*i.e.,* screenshot). However, there are important differences that make *screen parsing* applicable to a wider range of down-stream applications (Table 1).

*3.2.1 UI Element Detection.* UI Element Detection is a specific application of object detection, which extract a *set* of class-labeled bounding boxes from an image. When trained and applied to UI screens, the prediction output corresponds to the set of UI elements on the screen, which is useful on its own or as a "first-pass" step for further processing. The main difference from *screen parsing* is that UI Element results in a flat structure, which prevents it from representing relationships between elements. Heuristics can be applied to detect and group elements; however there is no guarantee that all elements will be connected.

Table 1: This table shows the requirements of several downstream applications and support for them among our implementation and related approaches. Screen parsing's problem formulation allows it to be applied more widely.

| Requirements | Complete | Grounded | Abstractive |
|---|---|---|---|
| **Applications** | | | |
| Structural Similarity | N | N | Y |
| Screen Reader | N | Y | N |
| Code Generation | Y | N | N |
| **Approaches** | | | |
| Scene Graph | Y | Y | N |
| GUI Skeleton | N | N | Y |
| Heuristics | Y | Y | N |
| Our Implementation | Y | Y | Y |

*3.2.2 GUI Skeleton Generation.* The GUI Skeleton is an artifact produced by the UI2Code system that describes the types of widgets in a screen and their hierarchical structure [6]. Similar to our model implementation, UI2Code is trained to produce trees processed from view hierarchies.

It is important to note that an app's GUI Skeleton is different its UI hierarchy (the target output of our model). Namely, it doesn't support we what refer to as *element grounding*, the ability to match items in its output to its input. For example, an app's GUI Skeleton might indicate that the screen contains a list container with three buttons, but it is unable to indicate which three buttons (on a screen with many buttons) belong to the list. Thus, the GUI Skeleton cannot be used to support certain applications, such as screen reader navigation.

*3.2.3 Scene Graph Generation.* Screen Graph Generation (SGG) is a visual scene understanding problem that models the relationships between visible objects using *scene graphs*. Like our model, SGG models are designed to process an input image and generate a graph whose nodes are detected objects in the scene and edges are semantic relationships between those objects.

Scene graphs are often constructed to describe real-world visual scenes [21]. Unlike UIs, which are typically constructed using nested views stemming from a single root node, visual scenes can contain multiple entities, represented as independent sub-graphs. We purposefully constrained our model to produce a single connected tree to reflect this property of UIs.

Most edges in a scene graph correspond to direct relationships between detected objects, and SGG models often consider *pairwise* relationships rather than *hierarchical* ones. Because of this, a strong and frequently-used baseline for SGG is computing the prior probabilities of relationships between object classes (ignoring position) on the training set [40]. Edges between leaf nodes are relatively rare among UI hierarchies, as most elements are indirectly joined by container elements.

## 4 IMPLEMENTATION

Our implementation of screen parsing uses separate models to *(i)* detect elements from a screenshot, *(ii)* group them together in a graph structure, and *(iii)* predict labels for the element groups.

### 4.1 UI Element Detection

We used a standard object detection model to extract the set of UI elements in a screen and their parameters. Specifically, we trained a Faster-RCNN [33] model with a ResNet-50 [16] backbone on our UI screen dataset. Before feeding an image to the element detection model, we resized images to 256x256 and normalized each input channel to have a mean of 0 and standard deviation of 0.5. We first run our detection model on an input screenshot and keep all detections that have a confidence of at least 0.7. We then apply non-max suppression to remove overlapping detections with lower confidence (IoU threshold of 0.5).

### 4.2 UI Hierarchy Prediction

After a set of detections is obtained from the Element Detection model, the next step is to predict their hierarchical relationship. A natural way of representing this is using a graph structure, where elements are linked to one another with parent-child relationships. Intuitively, the problem can be thought of as generating a complete graph (*i.e.,* the UI hierarchy) given the leaf nodes (*i.e.,* visible elements). We draw inspiration from the NLP literature on text parsing, where such graph structures are often used to define relationships between words in a sentence. Specifically, we build a top-down transition-based parser [25], which is able to construct any UI hierarchy[1], and offers fast and efficient decoding.

Like other transition-based parsers, our model incrementally produces a graph structure through a sequence of actions, and is most closely related to the approach detailed in similar dependency parsers used in NLP [25]. Our model uses three data structures to perform parsing: the input buffer ($l$) that holds the set of visible UI elements, the stack ($\sigma$) that allows the model to traverse the graph, and the set of visited nodes ($\alpha$). The actions that we support are:

- *Arc* – A directed edge is created between the node on top of $\sigma$ (parent) and the node in $l - \alpha$ with the highest attention score (child). The child is pushed onto $\sigma$ and added to $\alpha$.
- *Emit* – An intermediate node (represented as a zero-vector) is created and pushed onto $\sigma$.
- *Pop* – $\sigma$ is popped (*i.e.,* the top element is removed).

Figure 3 provides an example of how these actions are used to parse a screen.

*4.2.1 Model Architecture.* Our model architecture (Figure 3) consists of a LSTM-based encoder and decoder. Our chosen encoder model, the LSTM [17], is a type of recurrent neural network effective at encoding long sequences. LSTMs are designed with special gated memory cells that enable it to perform computations useful for our task, such as counting and recognizing hierarchy [15]. The input of the model is the list of UI elements in a screen, sorted using y-position as the primary key and x-position as the secondary key. Each element represented as the concatenation of its position and a one-hot class vector for the UI element type (*e.g.,* Text, Slider, Picture, etc...). The final hidden state is used as the initial state of the decoder.

---

[1] Some parsing algorithms are designed to handle only a subset of parse trees known as *projective trees*, which makes them difficult to apply to view hierarchies.

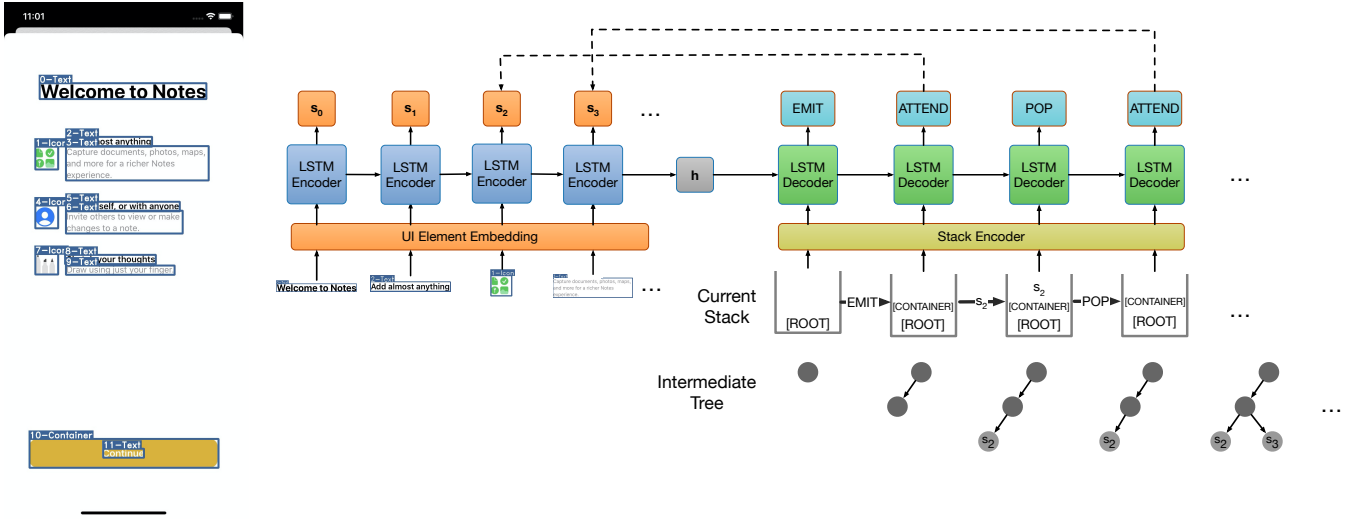

Figure 3: Our UI Hierarchy prediction model is a stack-based transition parser. A Bi-directional LSTM encoder is fed a set of embedded UI elements and query tokens. The final hidden state is used to initialize a LSTM decoder network. The decoder produces a sequence of actions that describe the UI hierarchy using a continuously updated state (stack, buffer, and visited set).

*4.2.2 Decoding.* At every decoding timestep, the LSTM is fed the *(i)* last hidden state ($h_t$) and *(ii)* the element at the top of $\sigma$. The LSTM returns *(i)* an output vector ($o_{t+1}$) and *(ii)* an updated hidden state ($h_{t+1}$). The output $o_{t+1}$ is fed through a linear layer that produces the logits for the emit and pop actions. The output $o_{t+1}$ is also used to compute the scaled dot-product attention between all of the encoded UI elements $\{s_0, s_1, ..., s_N\}$. Finally, an action vector is constructed by concatenating the emit ($u_e^i$) and pop ($u_p^i$) activations with the attention scores.

$$u_j^i = \frac{e_j^T h_i}{\sqrt{n}} \qquad (1)$$

$$p(a_i|a_0, a_1, ..., a_{i-1}, P) = softmax(concat(u_e^i, u_p^i, u^i)) \qquad (2)$$

$n$ is the size of the hidden state, $P$ represents the input and $a_0, a_1, ...a_n$ represent the previously selected actions. This process is repeated until all leaf nodes are added to $\alpha$, which guarantees that the generated graph is complete. Finally, as a heuristic to prevent repeated *Emit* and *Pop* actions, we set the probability of the *Emit* action to 0 if the last 10 actions does not contain an *Attend*.

The output of the model is additionally smoothed to remove extraneous intermediate nodes.

Table 2: Table of group labels considered for each dataset, along with number of occurrences.

| AMP | RICO |
| --- | --- |
| Tab Bar Button (63170) | List Item (56186) |
| Table (23693) | Toolbar (29068) |
| Tab Bar (19602) | Card (6091) |
| Collection (19420) | Drawer (5756) |
| Button (9779) | Multi-Tab (3189) |
| Segmented Control (2988) | Bottom Navigation (236) |

### 4.3 Group Labeling

To label the intermediate nodes in a tree, we train a separate classifier. We first inspect each dataset to determine the most common labels assigned to "containers" and select 7 classes (including an "Other" class) based on frequency and relevance to our task (Table 2).

Our Group Labeling classifier is based off the Deep Averaging Network (DAN) architecture used for sentence classification [18]. To classify a given node, we retrieve a list of all of its descendant elements. Each element in the list is embedded using using a feed-forward layer, and all of the embeddings are pooled using the sum operation. The pooled representation is fed into a MLP that predicts its label. Because some containers appear much more frequently than others, we use a weighted loss function for training (class-weighted cross entropy), and the F1-macro metric to measure validation and test performance. Our best group labeling models achieved F1-macro scores of 0.61 and 0.76, on AMP and RICO (our two training datasets).

This approach to classifying element groups is a simple one that does not model the joint probability of multiple element groups

(*e.g.,* the probability of one group's label conditioned on another's). We will improve this aspect of our system in future work.

## 5 TRAINING

In this section, we primarily describe the training procedure for our system's primary component – the UI Hierarchy model. We first describe how we extracted and processed a dataset for this purpose. Then, we describe an effective approach for training parsing models that is especially relevant to UI Hierarchy modeling.

### 5.1 Datasets

We trained our models on two mobile UI datasets: *(i)* **AMP**, an internal dataset of 130,000 iOS screens, and *(ii)* **RICO**, a publicly available dataset of 80,000 Android screens [8]. Each dataset contains screenshots, annotated screens, and their view hierarchies. Both datasets collected by crowdworkers who installed and explored popular apps across 20+ categories (in some cases excluding certain ones such as games, AR, and multimedia) on the iOS and Android app stores. More information is available in the original papers [8, 41]. Before training, three splits are created for each dataset: training (70%), validation (15%), and testing (15%). When training our system, we only train on screens with less than 64 elements (to make training more efficient), but we do not apply this constraint to our test set.

*5.1.1 Node Correspondence.* The first step is to match up visible elements with a corresponding node in the view hierarchy. We ran our trained UI Element detector on screenshots, which produced a list of detections above a confidence threshold (0.7). We employed a best-cost matching algorithm [22] to compute the best match between the set of element detections and the set of bounding boxes found in the view hierarchy. The matching score between two bounding boxes are defined as the intersection-over-union (IoU) score, and pairs with low scores (IoU < 0.5) are ignored.

*5.1.2 Extracting Hierarchical Information.* We found that many of the screens in our dataset had missing or mostly incomplete view hierarchies (*i.e.,* most of the visible elements did not have a corresponding element in the view hierarchy). For example, in the **AMP** dataset, we found that around 40,000 screens had view hierarchies that were suitable for ground truths. To train and evaluate our model on a higher-quality subset, filtered both datasets. The AMP dataset was filtered by selecting screens where at least 80% of annotated nodes had a corresponding element in the view hierarchy. The RICO dataset was filtered using scores from the node correspondence step – only screens where the average match score was greater than 0.8.

*5.1.3 Graph Smoothing.* Because the view hierarchy is an artifact of the UI rendering system, it contains some irrelevant nodes and edges that represent class inheritance and singleton containers. We preprocessed view hierarchy graphs using a smoothing algorithm that removed nodes which *(i)* only had one child and *(ii)* did not correspond to a visible element.

### 5.2 Training Algorithm

A standard approach to training transition-based parsers is defining an "oracle" function that produces a sequence of actions for every view hierarchy. An example of an oracle function for graph-structured data is running a depth-first search and recording the order nodes were entered and exited (Figure 4). We compared two different approaches to oracle training for our element grouping model.

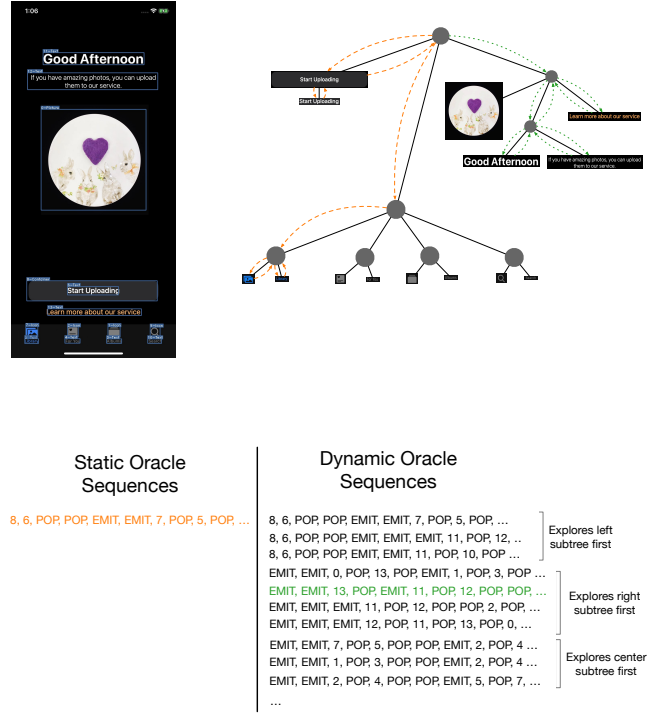

Figure 4: We explored two oracle-based training procedures for UI hierarchy prediction. For a given ground-truth UI hierarchy (top), a static oracle (left) only produces one sequence of optimal actions, while a dynamic oracle (right) produces all optimal sequences.

The first approach we compared is the *static oracle*, which is a simple and common implementation that traverses the graph deterministically (*i.e.,* produces exactly one sequence of actions for every graph). For screen parsing, this requires defining an ordering function that sets a deterministic order by which children are processed (*e.g.,* children are ordered top-down, left-to-right). During training, the parser is trained to maximize the likelihood of the static oracle's "gold" action at every timestep.

The second approach is a *dynamic oracle*, which provides a *set* of optimal actions at every state for the model to learn instead of a single action. During training, if the model's top-choice action is optimal then it is executed, and otherwise an optimal action from the oracle's output is randomly selected and executed. While other options are available [3, 13], we found that training to maximize the average likelihood of the set of optimal actions [26] led to the best results:

$$p(\mathbf{z_g}|\mathbf{p_t}) = \frac{1}{|\mathbf{z_g}|} \sum_{z_{g_i} \in \mathbf{z_g}} p(z_{g_i}|\mathbf{p_t}) \qquad (3)$$

# 6 EVALUATION

We compared our final system (Screen Parser Dynamic) to *(i)* a baseline system [41] and *(ii)* a baseline training procedure [25], and we show that our implementation significantly improves performance.

Screen Recognition is a heuristic-based system used to generate accessibility metadata from pixel data, and it is similar to heuristic-based approaches employed by other UI reverse-engineering work [29]. Similar to our system, Screen Recognition first runs an object detection model on a screenshot, which returns a set of element detections. These detections are then processed using a set of manually-defined heuristics that check for features such as nesting, text grouping, tab grouping, and picture subtitles.

We also used a baseline training procedure to train our system (Screen Parser Static), and we show that our chosen approach significantly outperforms standard training methods for NLP parsers.

To summarize, we compared the following systems in our evaluation:

- *RCNN Oracle* - This is not a system; it represents the best possible matching between the RCNN detections and the ground truth hierarchy. This gives a rough bound for the best-case parsing performance given the accuracy of the UI element detector.
- *Screen Recognition* - The complete Screen Recognition with its original UI element detector and heuristics.
- *Screen Recognition + RCNN* - The Screen Recognition heuristics run on the output of our RCNN-based UI element detector. When run on the RICO dataset, we used an RCNN model trained on the RICO dataset and mapped the the labels from the RICO label set to the AMP equivalent (the heuristics were designed for AMP).
- *Screen Parser Static* - The Screen Parser system where the UI Hierarchy model is trained using a static oracle (standard training procedure).
- *Screen Parser Dynamic* - The Screen Parser system where the UI Hierarchy model is trained using a dynamic oracle (improved training procedure).

## 6.1 Performance Metrics

To compare prediction outputs to ground truth view hierarchies, we first used our node correspondence algorithm (Section 5.1.1) to label the nodes in each graph with corresponding identifiers. Container nodes are matched using a similar method, where the score is the IoU of their descendant nodes.

We computed three types of metrics that measure different performance aspects relevant to down-stream tasks.

*6.1.1 Edge-based metrics.* Popular approaches to evaluating natural language parsers (*e.g.,* constituency parsing) are based on measuring the number of correctly predicted edges (*e.g.,* constituents) [19]. We decomposed both the ground truth and prediction graph into sets of edges and computed two metrics: *(i)* the overall F1 score and *(ii)* the F1 score for only edges that are attached to leaf nodes. The F1 score of the leaves can be more relevant for some downstream applications which use lower-level element groupings.

The F1 score is bounded from 0 to 1 and a higher score indicates better performance. If the prediction doesn't contain any matched nodes (possibly due to inaccurate element detection), the F1 scores for the overall tree and leaves are set to 0.

*6.1.2 Distance-based metrics.* While edge-based metrics are simple to compute, they can unfairly penalize some types of mistakes (*e.g.,* correct grouping but wrong parent). Graph Edit Distance (GED) is a measurement of graph similarity that considers the minimum number of "edits" needed to make a graph isomorphic to another.

$$GED(g_1, g_2) = \min_{(e_1,...,e_k) \in \mathcal{P}(g_1,g_2)} \sum_{i=1}^{k} c(e_i) \quad (4)$$

$\mathcal{P}(g_1, g_2)$ refers to the set of possible edit paths between $g_1$ and $g_2$. We consider GED that allows 4 edit operations all with cost of 1: the insertion and deletion of nodes and edges. Exact computation of GED is computationally expensive (NP-complete), so we use an inexact algorithm that approximates an upper bound of the true distance [11].

Because a lower GED indicates better performance, we set the GED to the number of edges in the ground truth tree if the prediction doesn't contain any matched nodes.

*6.1.3 Group-based metrics.* Finally, we considered group-based metrics that target the grouping of elements rather than their structure. This metric is more relevant for some downstream tasks such as screen segmentation that aim to partition the screen.

This metric is computed as the mean of each container's (*e.g.,* intermediate node) IoU score with the ground truth. Similar to edge-based metrics, the container match (CM) score is bounded between 0 and 1, where a score of 1 indicates that all groups were correctly matched. For trees without any matched modes, we set the score to 0.

## 6.2 Results

Table 3 shows the results of our performance evaluation using our set of metrics. Our results show that our final system, Screen Parser Dynamic outperforms all baselines in all performance metrics. In this section, we provide more detailed comparison with baselines and further analyze factors that impact performance.

*6.2.1 Comparison with Screen Recognition.* Both Screen Parser models outperform Screen Recognition on both datasets. One reason is that Screen Recognition and most other heuristics-based approach are not *abstractive*, which prevents them from producing "deep" trees. Performance on edges containing leaf nodes (*i.e.,* shallower relations) is generally much better; Compared to overall F1 score, Screen Recognition had a 25% higher F1 Leaves score.

More importantly, Screen Recognition was not designed to produce output similar to app view hierarchies; instead, it was designed to support common groupings required by screen reader navigation. In addition to the set of performance metrics described here, we recommend holistically evaluating systems in downstream tasks.

*6.2.2 Effect of Improved Training Procedure.* Based on our results, Screen Parser Dynamic performs up to 23% better than Screen Parser Static. Since both static and dynamic versions of our model was trained to maximize the likelihood of the same data, we can conclude that the dynamic oracle training technique is effective in increasing screen parsing performance.

Table 3: We evaluated screen parsing performance using 4 metrics: F1 score (F1), F1 score of edges with leaf nodes (F1 Leaves), graph edit distance (GED), and container match cost (CM). Higher is better for all metrics except GED. More details are described in the performance metrics section. Note that the RCNN Oracle is not a system – it is the best possible matching between the RCNN detections and the ground truth.

|  | AMP | | | | RICO | | | |
|---|---|---|---|---|---|---|---|---|
|  | F1↑ | F1 Leaves↑ | GED↓ | CM↑ | F1↑ | F1 Leaves↑ | GED↓ | CM↑ |
| RCNN Oracle | 0.76±0.22 | 0.75±0.22 | 16.6±20.9 | 0.79±0.19 | 0.89±0.14 | 0.89±0.14 | 8.8±15.9 | 0.93±0.07 |
| Screen Recognition | 0.40±0.20 | 0.52±0.26 | 23.5±20.7 | **0.63±0.23** | 0.39±0.19 | 0.47±0.26 | 23.8±21.4 | 0.43±0.19 |
| Screen Recognition + RCNN | 0.34±0.19 | 0.44±0.24 | 25.5±21.1 | 0.54±0.21 | 0.41±0.23 | 0.44±0.28 | 17.8±19.7 | 0.48±0.23 |
| Screen Parser Static | 0.53±0.23 | 0.62±0.22 | 26.1±24.6 | 0.59±0.16 | 0.61±0.27 | 0.59±0.27 | 15.2±16.2 | 0.69±0.24 |
| Screen Parser Dynamic | **0.60±0.23** | **0.67±0.23** | **20.2±20.9** | **0.63±0.16** | **0.66±0.28** | **0.64±0.28** | **13.2±15.5** | **0.74±0.24** |

Recall that the main difference between the two training procedures is that the static oracle only produces one sequence of optimal actions (*i.e.,* the canonical action sequence) while the dynamic oracle produces all optimal sequences (Figure 4). This is especially relevant for UI hierarchies, where the tree structure can be several levels high, leading to exponentially more possible optimal sequences. This is in contrast to natural language parse trees, which are typically limited by the relatively short length of sentences.

While the canonical action sequence provably correct (*i.e.,* contains all correct element relationships) [13], it leads to *exposure bias* – where the model is biased to perform well only in states it has seen during training. During test-time, the model may choose an action outside of this sequence (either by making an error or choosing another optimal action), which causes the model to perform poorly afterwards.

*6.2.3 Effect of UI Element Detection Performance.* All systems were fed UI element detections as input, and errors in the upstream model also affected the performance of the hierarchy prediction.

To estimate the upper-bound performance of systems that rely on the RCNN output, we included the *RCNN Oracle* which constructs an output using the best possible between the detector output and the ground truth hierarchy. Even with access to the ground truth, it does not achieve perfect accuracy – possibly a result of missing or inaccurate detections. This suggests that a better object detection model could further improve UI hierarchy prediction.

As an example, we ran Screen Recognition's heuristics on both its default object detector and our RCNN model's output. Compared to our system's RCNN model, Screen Recognition's object detector is optimized for the AMP dataset (*e.g.,* tuned per-class confidence threshold) which results in better performance.

*6.2.4 Performance on Complex Screens.* Finally, we analyzed the performance of screen parsing system on screens of different complexity. Figure 5 shows the overall F1 score for each system run on splits of the test data containing a screens with a specified # of elements. Performance is highest for screens up to 32 elements and degrades following that threshold. One major factor is lower object detection accuracy with smaller objects (screens with more elements tend to have smaller elements), since the performance of the RCNN Oracle also drops past that point. Interestingly, Screen

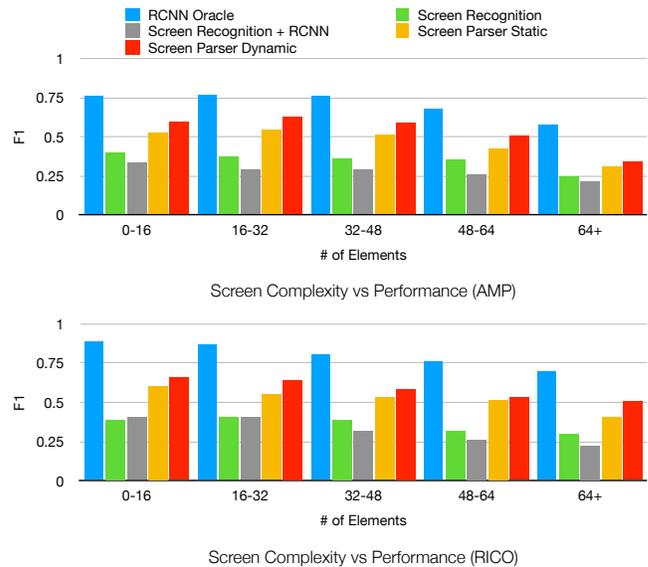

Figure 5: Analysis of each system's performance on screens of varying complexity. Screens with a higher number of elements introduce challenges for both UI element detection (screens with large # of elements generally have smaller and more dense elements) and UI hierarchy prediction.

Recognition's performance remains relatively constant, which suggests that many of the local patterns targeted by heuristics are not as affected by screen complexity. We also note that although both Screen Parser systems were only trained on screens with up to 64 elements, they still perform competitively for more complex screens.

Examples of failure cases (some of which result from these factors) are shown in Figure 6.

## 7 EXAMPLE APPLICATIONS

In this section, we present a suite of example applications implemented using our screen parsing model. These applications show the versatility of our approach and how the UI hierarchy predicted by our model can be used to facilitate many existing tasks.

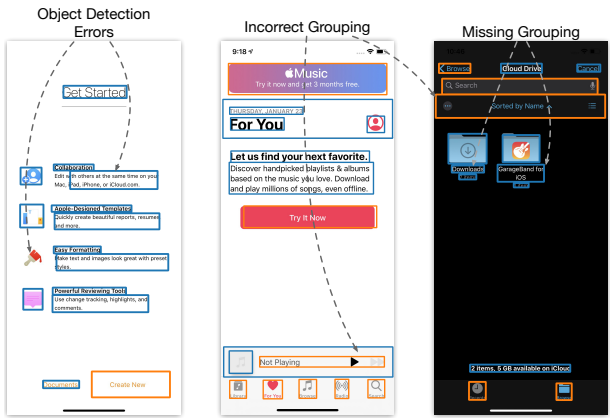

Figure 6: Examples of some errors by our screen parsing model. We identified three types of errors that can occur: *(i)* object detection errors, *(ii)* incorrect groupings, and *(iii)* missing groupings. Object detection errors can lead to missing elements or misaligned bounding boxes, which our model relies on to infer grouping. Incorrect groupings can assign irrelevant text labels to icons. Missing groupings can result in errors in downstream applications, such as a non-optimal navigation order for screen readers.

## 7.1 UI Similarity Search

Many recent efforts in modeling UIs have focused on representing them as fixed-length embedding vectors. These vectors can be trained to encode different properties of UI screens (*e.g.,* layout, content, and style) and support down-stream tasks. For example, a common application of embedding models is measuring screen similarity, which is represented by distance in embedding space. We believe the performance of such models can be improved by incorporating structural information, an important property of UIs.

Our implementation is trained to model the structural relationships between on-screen elements, and we show that its internal representations are useful for this purpose. To generate an embedding of a UI, we feed it into our model and pool the last hidden state of the encoder. This includes information about the position, type, and structure of on-screen elements. Figure 7 shows the 2-D projection [27] of randomly-sampled screens embedded using this technique. This set includes several variations of app screens, including *(i)* scaling, *(ii)* language, *(iii)* theme, and *(iv)* dynamic content. Our model is largely invariant to these changes, since their structure is the same, just rendered under different conditions. The properties of our embedding could be useful for some UI understanding applications, such as app crawling and information extraction where screens are characterized more by their semantics than appearance. We provide examples of UIs retrieved by our similarity search application in the appendix to illustrate the types of information our embedding captures.

## 7.2 Accessibility Enhancement

Screen readers help blind and visually impaired users access applications by reading out content highlighted by a cursor. Knowledge of UI element location (*i.e.,* spatial information) and hierarchy is important for screen readers to compute the correct order to move the cursor through the screen (*e.g.,* elements in the same group should be ordered consecutively), and for accessible apps, this information is found in an app's accessibility metadata. Recent work [41] has successfully generated missing metadata for inaccessible apps by running an object detection model on the UI screenshot. Their approach to generating hierarchical data relies on manually defined heuristics that detect and group localized patterns between elements (*e.g.,* a container with a text element inside it might be grouped as a button). However, these approaches may sometimes fail because they do not have access to global information that is necessary for resolving ambiguities.

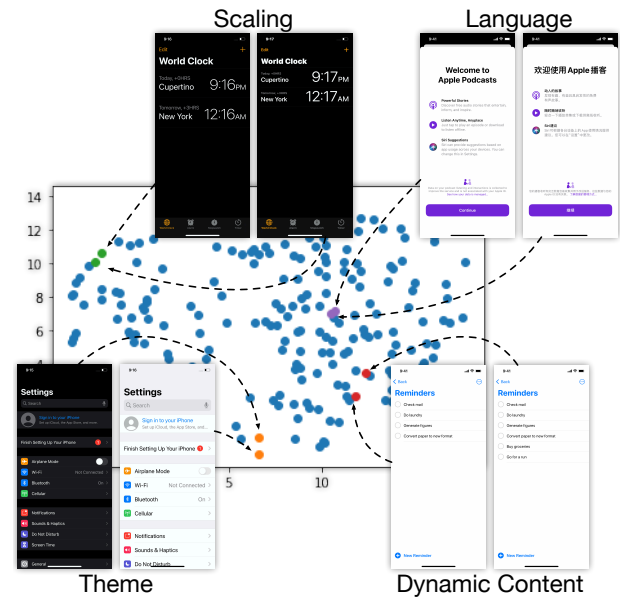

Figure 7: The intermediate representation of our parsing model can be used to produce a screen embedding, which describes hierarchical structure of an app. We embedded a set of app screens using our model and visualize them in a 2-D projection. We show that display settings such as *(i)* scaling, *(ii)* language, *(iii)* theme, and *(iv)* small dynamic changes result in minimal variation, which may be useful for some downstream tasks that rely on characterizing screens by semantic structure rather than aesthetic appearance.

In contrast, our implementation generates a UI hierarchy with a global view of the input, so it can overcome some of the limitations of heuristic-based approaches. We used the predicted UI hierarchy to group together the children of intermediate nodes of height 1 that contained at most one text label and used the X-Y cut algorithm [28] to determine navigation order. Figure 8 shows an example where the grouping output from the screen parser model is more accurate than the one produced by Screen Recognition heuristics. Note that this is not always the case. More examples are available in the appendix.

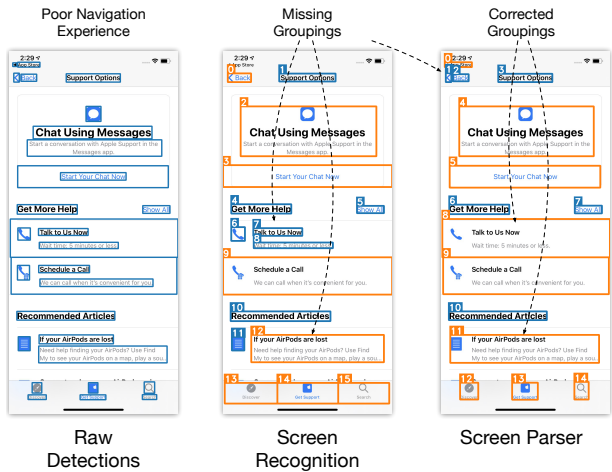

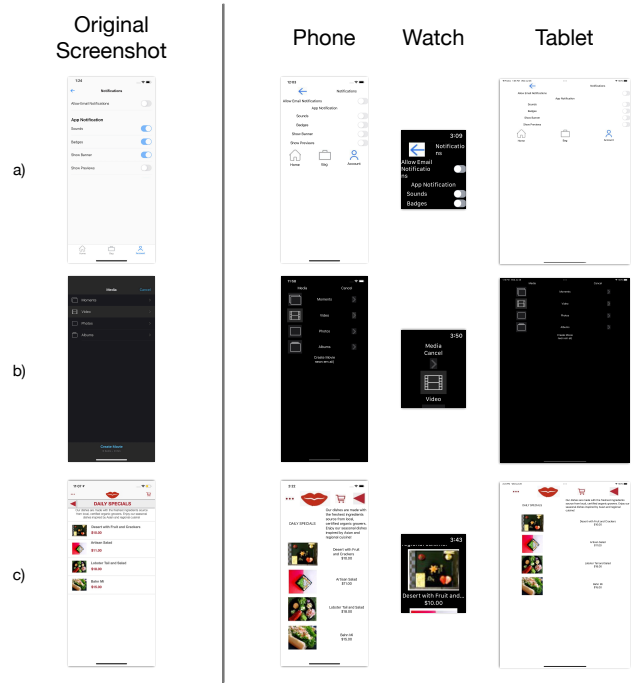

Figure 8: Recent approaches use object-detection approaches to generate accessibility metadata for inaccessible apps. Our model can be used to improve or augment the heuristic-based approach used by these systems to infer navigation order. Original detections from the object detector are shown in blue, and grouped elements are shown in orange. Element boxes are annotated using their navigation ordering [28], where the number represents how many swipes are needed to access the element when using a screen reader. While both results contain errors, in this case, Screen Parser correctly groups more elements, which decreases the number of swipes needed to access elements.

### 7.3 Generating UI Code from a Screenshot

Producing code from screenshots or mock-ups can greatly accelerate application prototyping development. A simple approach for code generation is *(i)* to first extract the location and type of UI elements using an object-detection model, *(ii)* then generate code that places the appropriate UI controls at the detected locations. While this approach may result in interfaces that are visually similar to the input, it is undesirable for several reasons. Code generated using this approach often uses absolute positioning constraints to instantiate UI controls, which prevents it from adapting to new screen sizes and makes it less useful for developers to work off of.

Some systems [29] use heuristics to detect a limited subset of containers (*e.g.,* lists), while others [5] augment visually detected elements with hierarchical data extracted from the window manager. To generate high quality, responsive code, structural understanding of a UI is an important step.

We built an example application that uses our implementation to generate SwiftUI code from a app screenshot. We employed a technique used by compilers to generate code from abstract syntax trees (AST) known as the *visitor pattern*. First a screenshot is fed into our system, which produces a UI hierarchy. We performed a depth-first traversal of the UI hierarchy using a *visitor* function that generates code based on the current state (current node and stack). Specifically, the visitor function emits a SwiftUI control (*e.g.,* Text, Toggle, Button) at every leaf node and emits a SwiftUI container

Figure 9: By mapping nodes in the UI hierarchy to declarative view-creation methods, we can generate code for a UI from its screenshot. Generating code from the hierarchy rather than the layout ensures that it is responsive across screen sizes, and we show the same output code rendered on different device form factors. Our example application may produce some errors due to missing style information (a, c) or inaccurate OCR (b).

(*e.g.,* VStack, HStack) at every intermediate node [2]. We manually created a mapping between nodes types in the screen parser tree and SwiftUI views and automatically required parameters such as label text using OCR. Elements containing graphics, such as image views or icons, are represented by an image patch cropped from the original screenshot, which are automatically included as assets. When generating code for small form-factors such as smartwatches, we replace horizontal containers with vertical ones due to limited space. Finally, our system uses a simple heuristic to determine whether the app uses a light or dark theme, and sets a preferred color scheme.

The resulting code describes the original UI using only relative constraints (even if the original UI was not), allowing it to act responsively to changes in screen size or device type (Figure 9). The generated code does not contain appearance and style information (*e.g.,* text size, element color), which is sometimes necessary to render a similar-looking screen. Nevertheless, prior work [6] has shown that such output can be a useful starting point for UI development, and we believe future work can improve upon our approach by detecting these properties.

---

[2]Information about SwiftUI controls and containers are available in the SwiftUI documentation: https://developer.apple.com/documentation/swiftui/views-and-controls

## 8 LIMITATIONS AND FUTURE WORK

In this paper, we presented the problem of *screen parsing* and implemented a baseline implementation that shows how structured information can be predicted from a UI's visual appearance. Specifically, our implementation predicts the *presentation model* from a UI's screenshot, for which we had a large dataset of examples (*i.e.*, view hierarchies) to facilitate machine learning. Some of our system's constraints (*e.g.*, can only produce directed trees) were purposefully introduced by us in service of our chosen target model.

We see multiple opportunities to improve our particular implementation. From our evaluation, we found that certain components, such as the UI Element Detector, can limit the performance of others that rely on it. The performance of our system can also be improved by modeling changes *e.g.*, incorporating visual information (*e.g.*, dominant color or visual embedding of an element) in our hierarchy prediction and improving our group labeling model. Some down-stream applications have different notions of performance. For example, when computing screen reader navigation, lower-level groupings (*i.e.*, close to leaf nodes) matters more. To more accurately assess our system's performance, we plan to evaluate it in the context of down-stream tasks. Our current model generates its output entirely from visual input (i.e., app screenshot), which minimizes its dependencies. Nevertheless, we also believe there is an opportunity to take advantage partial or incomplete view hierarchy, which our model can use in conjunction with visual information to improve performance [5].

More broadly, we hope to apply *screen parsing* to extract other types of structured semantics from UI screenshots (including those on other platforms such as web and desktop UIs), including those that describe data flow, interaction, and tasks. We expect that for some of these, we will be able to re-use much of our current architecture. Others might require adding or moving constraints (*e.g.*, predicting more general types of graphs that may include cycles). Furthermore, some types of models (*e.g.*, task models) might not be possible to infer from a single screenshot and would instead require a sequence of screens. Regardless, we expect the utility of automated UI systems to increase as they gain the ability to parse reason about structured semantics from UIs. We believe a promising application of screen parsing lies in tasks that require higher-level semantics such as task automation and programming-by-demonstration [24], which often require accessibility metadata to work.

## 9 CONCLUSION

In this paper, we introduced the problem of *screen parsing*, the prediction of structured UI models from visual information. In a comparison to three related problems, we show that our problem formulation and model is more suited to the unique properties of user interfaces. We described the architecture and training procedure for our reference implementation, which predicts an app's presentation model as a UI hierarchy with high accuracy, surpassing baseline algorithms and training procedures. In addition, we showed that the properties of our system allow it to simultaneously support a diverse array of down-stream applications: *(i)* UI similarity search, *(ii)* accessibility enhancement, and *(iii)* code generation from UI screenshots. More broadly, we believe our formulation of

*screen parsing* will allow automated systems to better reason about the underlying structure and purpose of UIs, facilitating more advanced and complex interactions.

## ACKNOWLEDGMENTS

We thank our reviewers for their feedback which helped improve this paper. This research was funded in part by a NSF GRFP Fellowship.

## A MODEL HYPERPARAMETERS

| Model | Hyperparameter | Value |
|---|---|---|
| Faster-RCNN | optimizer | SGD |
| | lr (base) | 0.01 |
| | lr (max) | 0.1 |
| Screen Parser | optimizer | Adam |
| | lr | 1e-4 |
| | weight decay | 1e-5 |
| | dropout | 0.25 |
| | hidden size | 256 |
| | hidden layers | 4 |
| Group Labeler | optimizer | Adam |
| | lr | 1e-4 |
| | weight decay | 1e-4 |
| | hidden size | 256 |
| | hidden layers | 1 |

All models were trained with early stopping that stopped training when validation loss did not improve for 10 epochs. We implemented our models using PyTorch [31] and PyTorch Lightning [10].

## B ORACLE PSEUDOCODE

```
for example in dataset:
    node = example.rootNode
    while not example.isTerminalNode(node):
        optimal = set()
        if len(node.children) == 0:
            # at a leaf node, go back up
            optimal.add(PopAction())
        else:
            for child in node.children:
                if child.isLeaf:
                    optimal.add(ArcAction(child))
                else:
                    optimal.add(EmitAction())
        if dynamicTraining:
            model.train(optimal)
            if model.highestScoring in optimal:
                action = model.prediction
            else:
                action = randomChoice(optimal)
        else: # static training
            action = getCanonicalAction(optimal)
            model.train(action)
        node = node.nodeAfterAction(action)
```

## C UI RETRIEVAL EXAMPLES

Figure 10 shows examples of UIs retrieved using our UI Similarity Search example application. We embedded a set of query UIs and used them to retrieve similar UIs from a subset of our AMP dataset.

## D ACCESSIBILITY ENHANCEMENT EXAMPLES

Figure 11 shows examples of accessibility metadata generated by our system and Screen Recognition, a baseline that we compared with. Both systems occasionally produce minor errors (*e.g.,* grouping elements that do not belong together) but significantly improve the navigation experience.

## REFERENCES

[1] Airbnb. 2017. Sketching Interfaces. https://airbnb.design/sketching-interfaces/
[2] AutoIt. 2021. Function PixelSearch. https://www.autoitscript.com/autoit3/docs/functions/PixelSearch.htm
[3] Miguel Ballesteros, Yoav Goldberg, Chris Dyer, and Noah A Smith. 2016. Training with exploration improves a greedy stack-LSTM parser. *arXiv preprint arXiv:1603.03793* (2016).
[4] Tony Beltramelli. 2018. pix2code: Generating code from a graphical user interface screenshot. In *Proceedings of the ACM SIGCHI Symposium on Engineering Interactive Computing Systems*. 1–6.
[5] Tsung-Hsiang Chang, Tom Yeh, and Rob Miller. 2011. Associating the visual representation of user interfaces with their internal structures and metadata. In *Proceedings of the 24th annual ACM symposium on User interface software and technology*. 245–256.
[6] Chunyang Chen, Ting Su, Guozhu Meng, Zhenchang Xing, and Yang Liu. 2018. From UI design image to GUI skeleton: a neural machine translator to bootstrap mobile GUI implementation. In *Proceedings of the 40th International Conference on Software Engineering*. 665–676.
[7] Jieshan Chen, Chunyang Chen, Zhenchang Xing, Xiwei Xu, Liming Zhut, Guoqiang Li, and Jinshui Wang. 2020. Unblind your apps: Predicting natural-language labels for mobile GUI components by deep learning. In *2020 IEEE/ACM 42nd International Conference on Software Engineering (ICSE)*. IEEE, 322–334.
[8] Biplab Deka, Zifeng Huang, Chad Franzen, Joshua Hibschman, Daniel Afergan, Yang Li, Jeffrey Nichols, and Ranjitha Kumar. 2017. Rico: A mobile app dataset for building data-driven design applications. In *Proceedings of the 30th Annual ACM Symposium on User Interface Software and Technology*. 845–854.
[9] Morgan Dixon and James Fogarty. 2010. Prefab: implementing advanced behaviors using pixel-based reverse engineering of interface structure. In *Proceedings of the SIGCHI Conference on Human Factors in Computing Systems*. 1525–1534.
[10] WA Falcon and .al. 2019. PyTorch Lightning. *GitHub. Note: https://github.com/PyTorchLightning/pytorch-lightning* 3 (2019).
[11] Andreas Fischer, Kaspar Riesen, and Horst Bunke. 2017. Improved quadratic time approximation of graph edit distance by combining Hausdorff matching and greedy assignment. *Pattern Recognition Letters* 87 (2017), 55–62.
[12] Krzysztof Gajos and Daniel S Weld. 2004. SUPPLE: automatically generating user interfaces. In *Proceedings of the 9th international conference on Intelligent user interfaces*. 93–100.
[13] Yoav Goldberg and Joakim Nivre. 2012. A dynamic oracle for arc-eager dependency parsing. In *Proceedings of COLING 2012*. 959–976.
[14] W3C Working Group. 2014. Introduction to Model-Based User Interfaces. https://www.w3.org/TR/mbui-intro/
[15] Michael Hahn. 2020. Theoretical limitations of self-attention in neural sequence models. *Transactions of the Association for Computational Linguistics* 8 (2020), 156–171.
[16] Kaiming He, Xiangyu Zhang, Shaoqing Ren, and Jian Sun. 2016. Deep residual learning for image recognition. In *Proceedings of the IEEE conference on computer vision and pattern recognition*. 770–778.
[17] Sepp Hochreiter and Jürgen Schmidhuber. 1997. Long short-term memory. *Neural computation* 9, 8 (1997), 1735–1780.
[18] Mohit Iyyer, Varun Manjunatha, Jordan Boyd-Graber, and Hal Daumé III. 2015. Deep unordered composition rivals syntactic methods for text classification. In *Proceedings of the 53rd annual meeting of the association for computational linguistics and the 7th international joint conference on natural language processing (volume 1: Long papers)*. 1681–1691.
[19] Dan Jurafsky and James H. Martin. 2020. *Speech & language processing 3rd ed. draft*.
[20] Boris Knyazev, Harm de Vries, Cătălina Cangea, Graham W Taylor, Aaron Courville, and Eugene Belilovsky. 2020. Graph density-aware losses for novel compositions in scene graph generation. *arXiv preprint arXiv:2005.08230* (2020).
[21] Ranjay Krishna, Yuke Zhu, Oliver Groth, Justin Johnson, Kenji Hata, Joshua Kravitz, Stephanie Chen, Yannis Kalantidis, Li-Jia Li, David A Shamma, et al. 2017. Visual genome: Connecting language and vision using crowdsourced dense image annotations. *International journal of computer vision* 123, 1 (2017), 32–73.
[22] Harold W Kuhn. 1955. The Hungarian method for the assignment problem. *Naval research logistics quarterly* 2, 1-2 (1955), 83–97.
[23] James A Landay. 1996. SILK: sketching interfaces like krazy. In *Conference companion on Human factors in computing systems*. 398–399.
[24] Toby Jia-Jun Li, Amos Azaria, and Brad A Myers. 2017. SUGILITE: creating multimodal smartphone automation by demonstration. In *Proceedings of the 2017 CHI conference on human factors in computing systems*. 6038–6049.
[25] Xuezhe Ma, Zecong Hu, Jingzhou Liu, Nanyun Peng, Graham Neubig, and Eduard Hovy. 2018. Stack-pointer networks for dependency parsing. *arXiv preprint arXiv:1805.01087* (2018).
[26] Dhruv Mahajan, Ross Girshick, Vignesh Ramanathan, Kaiming He, Manohar Paluri, Yixuan Li, Ashwin Bharambe, and Laurens Van Der Maaten. 2018. Exploring the limits of weakly supervised pretraining. In *Proceedings of the European Conference on Computer Vision (ECCV)*. 181–196.
[27] Leland McInnes, John Healy, and James Melville. 2018. Umap: Uniform manifold approximation and projection for dimension reduction. *arXiv preprint arXiv:1802.03426* (2018).
[28] J-L Meunier. 2005. Optimized xy-cut for determining a page reading order. In *Eighth International Conference on Document Analysis and Recognition (ICDAR'05)*. IEEE, 347–351.
[29] Tuan Anh Nguyen and Christoph Csallner. 2015. Reverse engineering mobile application user interfaces with remaui (t). In *2015 30th IEEE/ACM International Conference on Automated Software Engineering (ASE)*. IEEE, 248–259.
[30] Jeffrey Nichols, Brad A Myers, Michael Higgins, Joseph Hughes, Thomas K Harris, Roni Rosenfeld, and Mathilde Pignol. 2002. Generating remote control interfaces for complex appliances. In *Proceedings of the 15th annual ACM symposium on User interface software and technology*. 161–170.
[31] Adam Paszke, Sam Gross, Soumith Chintala, Gregory Chanan, Edward Yang, Zachary DeVito, Zeming Lin, Alban Desmaison, Luca Antiga, and Adam Lerer. 2017. Automatic differentiation in pytorch. (2017).
[32] Angel R Puerta. 1997. A model-based interface development environment. *IEEE Software* 14, 4 (1997), 40–47.
[33] Shaoqing Ren, Kaiming He, Ross Girshick, and Jian Sun. 2015. Faster r-cnn: Towards real-time object detection with region proposal networks. *arXiv preprint arXiv:1506.01497* (2015).
[34] Richard S. Schwerdtfeger. 1991. Making the GUI Talk. ftp://service.boulder.ibm.com/sns/sr-os2/sr2doc/guitalk.txt
[35] Michael Shilman, Percy Liang, and Paul Viola. 2005. Learning nongenerative grammatical models for document analysis. In *Tenth IEEE International Conference on Computer Vision (ICCV'05) Volume 1*, Vol. 2. IEEE, 962–969.
[36] Richard Socher, Cliff Chiung-Yu Lin, Andrew Y Ng, and Christopher D Manning. 2011. Parsing natural scenes and natural language with recursive neural networks. In *ICML*.
[37] Zhuowen Tu, Xiangrong Chen, Alan L Yuille, and Song-Chun Zhu. 2005. Image parsing: Unifying segmentation, detection, and recognition. *International Journal of computer vision* 63, 2 (2005), 113–140.
[38] Jianwei Yang, Jiasen Lu, Stefan Lee, Dhruv Batra, and Devi Parikh. 2018. Graph r-cnn for scene graph generation. In *Proceedings of the European conference on computer vision (ECCV)*. 670–685.
[39] Tom Yeh, Tsung-Hsiang Chang, and Robert C Miller. 2009. Sikuli: using GUI screenshots for search and automation. In *Proceedings of the 22nd annual ACM symposium on User interface software and technology*. 183–192.
[40] Rowan Zellers, Mark Yatskar, Sam Thomson, and Yejin Choi. 2018. Neural motifs: Scene graph parsing with global context. In *Proceedings of the IEEE Conference on Computer Vision and Pattern Recognition*. 5831–5840.
[41] Xiaoyi Zhang, Lilian de Greef, Amanda Swearngin, Samuel White, Kyle Murray, Lisa Yu, Qi Shan, Jeffrey Nichols, Jason Wu, Chris Fleizach, et al. 2021. Screen Recognition: Creating Accessibility Metadata for Mobile Applications from Pixels. *arXiv preprint arXiv:2101.04893* (2021).
[42] Song-Chun Zhu and David Mumford. 2007. *A stochastic grammar of images*. Now Publishers Inc.

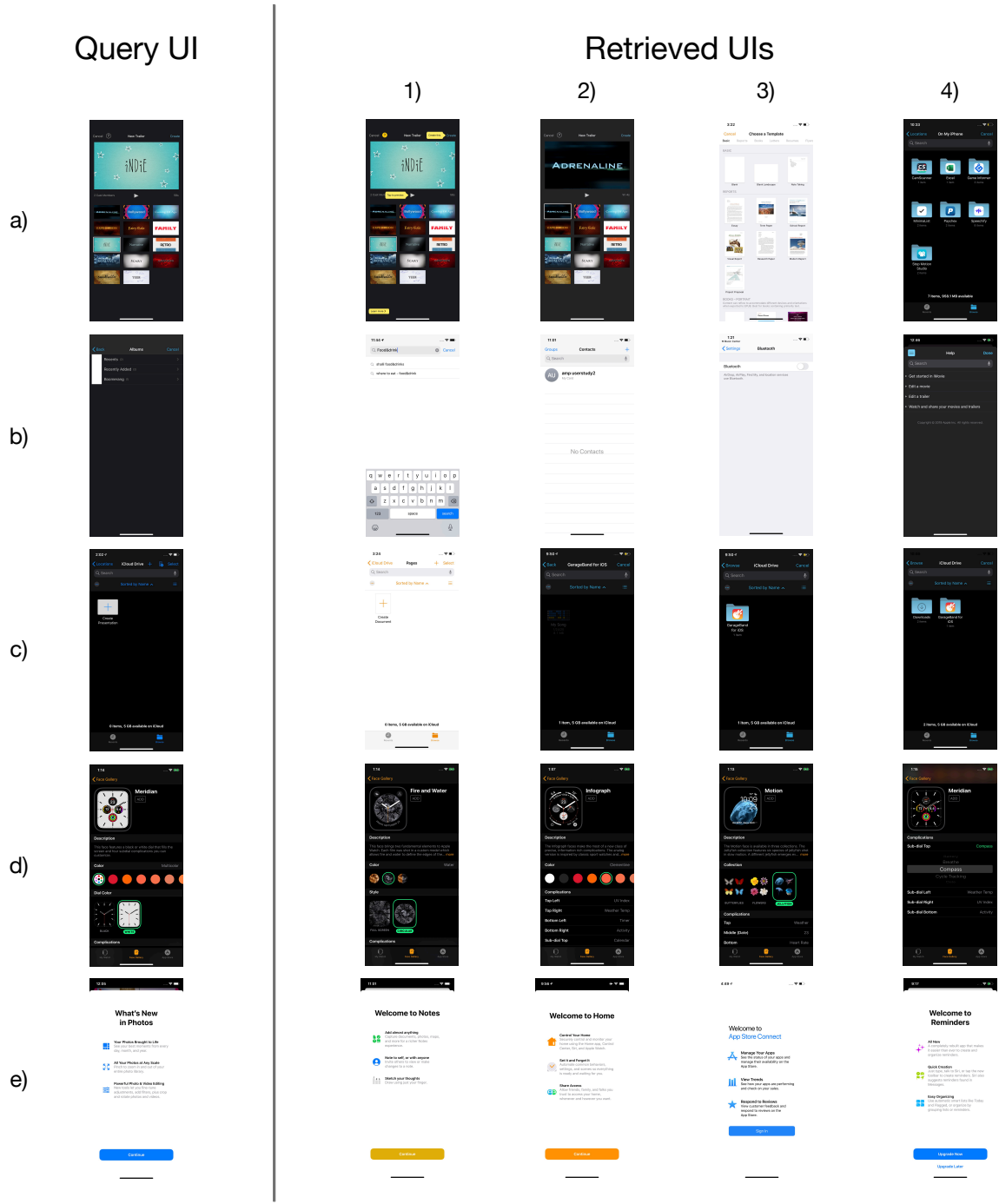

Figure 10: Example output of our UI Similarity Search example applications. We use several query UIs to find similar UIs in a subset of the AMP dataset. Retrieved UIs are ordered by their similarity to the query UI in embedding space. Many of the retrieved screens are from other apps with similar structural layout.

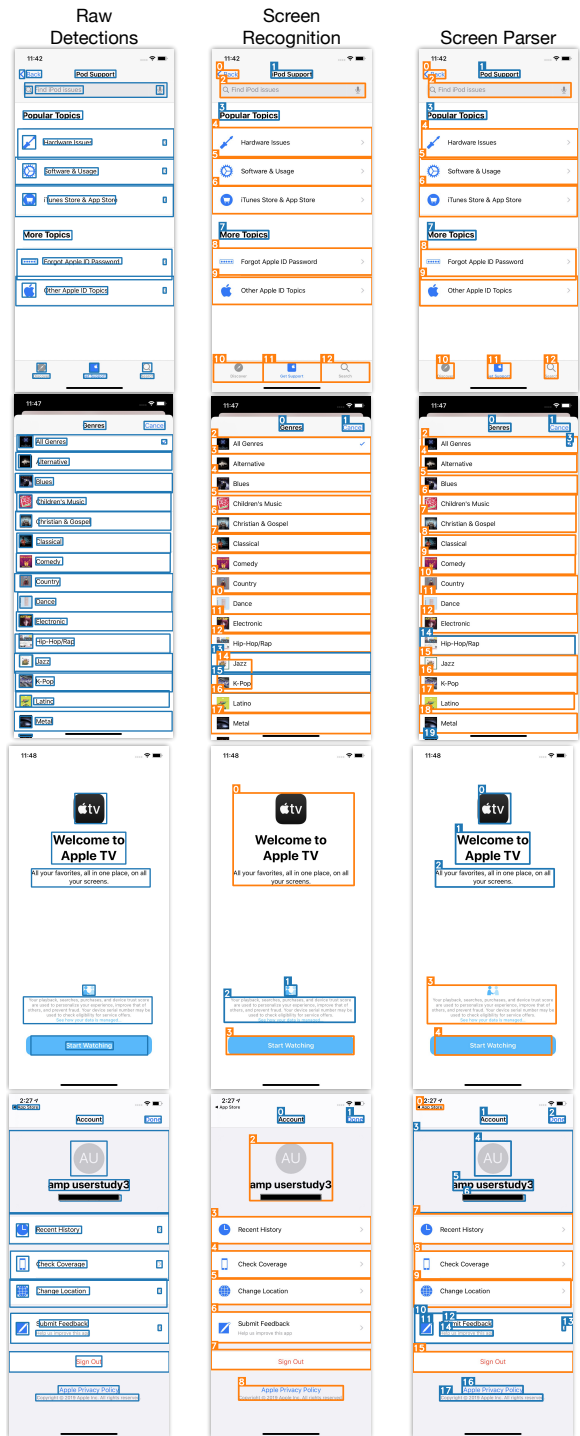

Figure 11: Examples of accessibility metadata generated for raw detections by Screen Recognition heuristics and our screen parser model. Each element is annotated with the number of swipes needed to reach it using a screen reader. Elements groups are shown in orange. The last row of screenshots contain an email address, which is redacted.